\documentclass[prl,twocolumn,showpacs,amsmath,superscriptaddress]{revtex4}

\usepackage{color,graphicx}
\usepackage{wasysym}
\usepackage{times}
\newcommand{\Svec}{\mathbf{S}}
\newcommand{\Qvec}{\mathbf{Q}}
\newcommand{\PP}{\mathcal{P}}
\newcommand{\MF}{\mathrm{MF}}
\newcommand{\eff}{\mathrm{eff}}

\newcommand{\tr}{\mathrm{Tr}}
\newcommand{\half}{\frac{1}{2}}

\begin{document}

\title{Semiclassical ordering in the large-$N$ pyrochlore antiferromagnet}
\author{U.~Hizi} 
\email{uh22@cornell.edu}
\affiliation{Laboratory of Atomic and Solid State Physics,
     Cornell University, Ithaca, NY, 14853-2501}
\author{Prashant~Sharma}
\affiliation{Laboratory of Atomic and Solid State Physics,
     Cornell University, Ithaca, NY, 14853-2501}
\affiliation{Argonne National Laboratory, Materials Science Division, Argonne, IL, 60439}
\author{C.~L.~Henley}
\affiliation{Laboratory of Atomic and Solid State Physics,
     Cornell University, Ithaca, NY, 14853-2501}

\begin{abstract}
We study the semiclassical limit of the $Sp(N)$ generalization of the
pyrochlore lattice Heisenberg antiferromagnet
by expanding about the $N\!\to\!\infty$ saddlepoint in powers of a generalized
inverse spin.
To leading order, we write down an effective Hamiltonian as a series in loops
on the lattice.
Using this as a formula for calculating the energy of any classical 
ground state, we
perform Monte Carlo simulations and find a unique collinear ground state.
This state is not a ground state of linear spin-wave theory,
and can therefore not be a physical ($N=1$) semiclassical ground state.
\end{abstract}

\maketitle

Geometrically frustrated antiferromagnets~\cite{ramirez_review,diep}
have attracted interest because their large classical ground state
degeneracy can allow a rich variety of correlated states, 
including (at $T=0$) quantum spin liquids or complex ordered states. 
The simplest examples are nearest-neighbor,
exchange-coupled antiferromagnets
in which spins from triangles or tetrahedra that share corners~\cite{clh_heff}:
the kagom\'e, ``checkerboard'', and
SCGO ($\mathrm{SrCr_{9p}Ga_{12-9p}O_{19}}$)
lattices~\cite{ramirez_review}
in two dimensions, plus the garnet and pyrochlore lattices
in three dimensions: the pyrochlore, in particular, 
consists of tetrahedra whose centers form a diamond lattice.
The Hamiltonian is
$H = \sum J_{ij} \Svec_i \cdot \Svec_j$,
where $J_{ij}\!=\!1$ for nearest neighbors $\langle i j \rangle$.
In fact, additional terms --
dipole interactions and anisotropies
(as in $\mathrm{Gd}_2\mathrm{Ti}_2\mathrm{O}_7$),
magnetoelastic couplings (as in
$\mathrm{ZnV}_2\mathrm{O}_4$ and $\mathrm{ZnCr}_2\mathrm{O}_4$)--
decide the order in most real materials~\cite{ramirez_review,diep}.
Still, the case with pure Heisenberg exchange
is worth understanding since 
(i) most simulations are done for this case; (ii) the more
realistic systems emerge from it by the addition of perturbations;
(iii) this has motivated experimentalists to search for model systems in which 
the aforementioned perturbations are small;
(iv) quantum effects can be studied without being overshadowed by classical
effects.

What is the ground state for large spin length $S$?
In unfrustrated antiferromagnets, it is just the classical
ground state dressed with zero-point fluctuations of
harmonic spin waves, and in frustrated cases the spin-wave 
zero-point energy \emph{may} lift the degeneracy of classical 
ground states~\cite{shender+clh}.
In the pyrochlore case, though, a large degeneracy remains~\cite{clh_harmonic};
its resolution by higher-order (anharmonic) terms in the semiclassical
($1/S$) expansion requires arduous approximations~\cite{uh_quartic}.

An established alternative to the spin-wave approach is 
to generalize the Heisenberg spins
[with SU($2$) $\cong$ Sp($1$) symmetry] to Sp($N$)
symmetry~\cite{read+sachdev_LN}:
here $N$ is the number of flavors of Schwinger 
bosons whose bilinear form represents 
a \emph{generalized spin}~\cite{read+sachdev_LN}, 
with length $\kappa=2S$.
The resulting mean-field theory (valid in the
$N\to\infty$ limit) is popular as an analytic
approach to the $S=1/2$ limit,
since the \emph{small}-$\kappa$ limit captures various disordered and exotic 
ground states~\cite{read+sachdev_LN,bernier_SL}.
The large-$N$ mean-field theory is also useful
at \emph{large}-$\kappa$ for this paper's problem, 
since it gives a simple analytical prescription for ground state
selection: unlike the spin-wave expansion,
here all degeneracies are (typically) broken at the
lowest order [$\mathcal{O}(1/\kappa)]$ quantum
correction~\cite{sachdev_kagome,read+sachdev_LN}.

However, on highly frustrated lattices this approach has the complication of
a macroscopic (exponential) number of degenerate saddle-points not
related by symmetry, so it is unknown a priori which of these should
be expanded around; this was handled till now by limiting the investigation
to ordering patterns of high symmetry and small magnetic cells, 
or by enumerating all saddle-points in a small finite 
system~\cite{sachdev_kagome,bernier}. 

In this letter, 
we develop an \emph{effective Hamiltonian}~\cite{clh_heff}
approach to this question.
The pertinent saddle-points are labeled by arrangements of valence bond
variables,
and we obtain a simple formula for the large-$N$
mean-field energy of \emph{any}
classical ground state, as a function of these variables.
The effective Hamiltonian 
is constructed as an analytical real-space expansion of 
\emph{loops} made of valence bonds.
This allows us to systematically search for a collinear pyrochlore ground state,
using Monte Carlo annealing, on quite large system sizes.
However, we also find that the pyrochlore ground state does \emph{not}
agree with even the lowest-order term in the spin-wave expansion, 
and therefore cannot give the right answer 
for the physical ($N=1$) ground state, in the large-$S$ limit, 
demonstrating a limitation of the large-$N$ approach for this case.

\emph{Large $N$ mean field theory.}\textemdash
We begin by discussing the mean-field Hamiltonian derived from the Sp($N$) generalization of $H$. For the  $N=1$ case we can write the spin interaction in terms of Schwinger boson operators as 
$\vec S_i\cdot\vec S_j=b^\dagger_{i\sigma}b_{i\sigma'}b^\dagger_{j\sigma'}b_{j\sigma}$,
where a sum over repeated indices $\sigma$ and $\sigma'$
(that take values $\uparrow,\downarrow$) is implied.
The Hilbert space of the spin model is obtained by constraining the number
of bosons on each site $b^\dagger_{i\sigma}b_{i\sigma}=2S$.
We can rewrite the interaction in terms of valence bonds created by the
operator $\epsilon_{\sigma\sigma'}b^\dagger_{i\sigma}b^\dagger_{j\sigma'}$,
where $\epsilon_{\uparrow\downarrow}=-\epsilon_{\downarrow\uparrow}=1$.
An arbitrary singlet state can be written in terms of some arrangement of these
bonds with at most $2S$ bonds emanating from any lattice site.
Generalizing these bond operators to $N$-flavors allows us to put a large number
of bonds on a link.
Since the Hamiltonian acting on a state changes at most two bonds per link,
the relative change in the number of bonds goes like $1/NS$.
In the large-$N$ limit, their fluctuations are quenched.
Therefore, we factorize the interaction in terms of valence bonds 
$Q_{ij}=\langle\epsilon_{\sigma\sigma'}b_{i\sigma,m}b_{j\sigma',m}\rangle/N$,
where the \emph{flavor} index $m=1,2...,N$.
We treat $Q_{ij}$ as classical quantities, to obtain the mean-field Hamiltonian
\begin{eqnarray}
H_\MF&=& \frac{1}{2}\sum_{\langle ij\rangle}\left[ N|Q_{ij}|^2 +
\left(\epsilon_{\sigma\sigma'}b^\dagger_{i\sigma,m}b^\dagger_{j\sigma',m} Q_{ij}
+H.c.\right)\right]\nonumber\\
&+&\sum_i\lambda_i\left(b^\dagger_{i\sigma,m}b_{i\sigma,m}-N\kappa\right)
\label{eq:h_LN}
\end{eqnarray}
Here Lagrange multipliers $\lambda_i$ have been introduced to enforce the
constraint on boson number $N \kappa$ at every site $i$,
defining the \emph{generalized spin length} $\kappa\!=\!2S$.
In what follows, we shall take $\lambda_i$
to be spatially uniform $\lambda_i=\lambda$.
We are interested in large enough values of $\kappa$ 
to condense a flavor mode of the itinerant bosons,
$\langle b_{i\sigma,m}\rangle=\sqrt{N}\delta_{1,m} x_{i\sigma}$,
for long-range order to develop.
The mean-field ground state energy (per flavor) is obtained by
diagonalizing~(\ref{eq:h_LN}) by a canonical Bogoliubov transformation: 
\begin{subequations}
\begin{eqnarray}
\frac{E_\MF}{N}&=&\frac{1}{2}\sum_{\langle ij\rangle}\left[|Q_{ij}|^2
+\left(\epsilon_{\sigma\sigma'}x_{i\sigma}x_{j\sigma'} Q^*_{ij}+
c.c.\right)\right]
\nonumber \\
&+& \sum_i \lambda\left(|x_{i\sigma}|^2-\kappa\right)
\label{eq:eclassical}\\
&+&\frac{1}{2}[\tr{\sqrt{\lambda^2\openone-\Qvec^\dagger \Qvec}}-N_s\lambda]
\label{eq:equantum}
\end{eqnarray}
\end{subequations}
Here $N_s$ is the number of lattice sites, and (\ref{eq:equantum}) is the 
zero-point energy contribution of the bosons.
The exact mean-field ground state is obtained by a constrained minimization of
the above expression.
It can be systematically approached as an expansion in powers of $1/\kappa$.
The leading contribution to the energy (of order $\kappa^2$) comes from terms
in (\ref{eq:eclassical}),
whose minimization simply relates the valence bonds to
the condensate configuration in the classical ground state(s) of the Heisenberg
Hamiltonian $H$ with spin size $\kappa/2$.
We will denote this configuration  of bond variables with a 
superscript $c$: $\{Q^c_{ij}\}$.
The quantum correction (of order $\kappa$) is provided by terms 
in~(\ref{eq:equantum}) for these bond configurations.

The ground states of the classical Hamiltonian~(\ref{eq:eclassical})
consist of all spin configurations in
which the spin vectors sum to zero in every tetrahedron.
On general grounds we expect quantum corrections to select \emph{collinear}
ground states from the classical
manifold~\cite{shender+clh,clh_heff,uh_harmonic}.
We therefore restrict our attention to such states, in which each spin can 
be denoted by an Ising variable $\eta_i\! \in\! \{\pm 1\}$. 
Collinearity implies that, up to an
arbitrary gauge transformation, $Q^c_{ij}=\kappa (\eta_i - \eta_j)/2$
and thus the bond variables are $\pm\kappa$ for every satisfied,
antiferromagnetic (AFM) bond, and zero otherwise.
Also, $\lambda^c\! =\! 4 \kappa $
for all pyrochlore lattice classical ground states.

\emph{Loop expansion and effective Hamiltonian.}\textemdash
Next,we recast the first quantum correction to the mean field
energy, Eq.~(\ref{eq:equantum}), for a given collinear
classical ground state, into an effective Hamiltonian
form where only some of the degrees of freedom remain~\cite{clh_heff}.
Eq.~(\ref{eq:equantum}) can formally be Taylor-expanded 
\begin{equation}\label{eq:expand}
\frac{E_q}{N}= -\half \sum_{m=1}^\infty
\frac{(2m+1)!!}{2^m \lambda^{2m-1} m!} \tr \left( \Qvec^\dagger \Qvec \right)^m
\end{equation}
Since $|Q_{ij}/\kappa|=1$ for AFM bonds, and zero
otherwise, $\tr (\Qvec^\dagger \Qvec/\kappa^2)^m$ is equal to the number 
of closed paths of length $2m$, composed of AFM bonds.
All terms in Eq.~(\ref{eq:expand}) depend solely on the structure
of the network formed by AFM bonds.
Note that since this network is bipartite, each nonzero element of
$\Qvec^\dagger \Qvec$ is $\kappa^2$.

\begin{figure}
\resizebox{!}{3.6cm}{\includegraphics{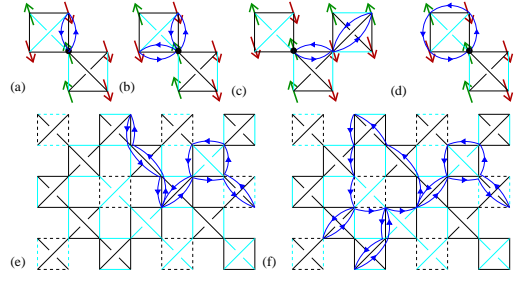}}
\caption{\label{fig:paths}
(a)-(d) 
Schematic diagram of terms contributing to the constant
term in the effective energy, due to $\tr \Qvec^2$ (a), and $\tr \Qvec^4$
(b,c,d). These are  (001) projections, where
the crossed squares are projected tetrahedra, and
AFM bonds are shown in dark.
All paths that do not contain loops, e.g. (a,b,c), 
can be viewed as paths on a coordination $4$ Bethe lattice.
(e)-(f):
Examples of the two types of paths that we need to count, in order to calculate
the effective Hamiltonian coefficients, as shown on a (001) slice
of the pyrochlore lattice. The dashed lines represent 
bonds that connect to adjacent slices.
(e) A decorated Bethe lattice path of length $14$ contributing to $F(14)$ and 
(f) A path of length $22$ containing a loop of length $8$, contributing
to $G(8,14)$.}
\end{figure}

In any collinear classical ground state, each tetrahedron has two up
spins and two down spins, and four AFM bonds forming a closed loop
(see Figs.~\ref{fig:paths}a-d).
This means that the local connectivity of the AFM network is identical for
all states, and many closed paths only contribute state-independent terms to 
Eq.~(\ref{eq:expand}).
For example, $\tr \Qvec^\dagger \Qvec=4N_s \kappa^2$,
for any classical ground state
since the only paths of length $2$ involve 
going to and fro on the same bond, and each site has four neighbors which 
have the opposite spin (see Fig.~\ref{fig:paths}a).
Similarly $\tr (\Qvec^\dagger \Qvec)^2= (16+12+4)N_s \kappa^4$,
where the three terms correspond to the 
paths shown in Figs.~\ref{fig:paths}b,~\ref{fig:paths}c,~\ref{fig:paths}d,
respectively.
All paths that do not involve loops, (e.g. those in 
Figs.~\ref{fig:paths}a,~\ref{fig:paths}b,~\ref{fig:paths}c)
can be viewed as paths on a Bethe lattice of coordination $4$,
and would contribute a constant term to the energy for all collinear classical
ground states.
The same is true for paths involving only trivial loops, in addition to the
Bethe lattice path, as in Fig.~\ref{fig:paths}d.
Here, a ``trivial'' loop is the loop of length $4$ that exists
within any tetrahedron.
The lowest order terms in expansion~(\ref{eq:expand})
that contribute a state-dependant 
term in the effective Hamiltonian are for $2m=6$, since the shortest
non-trivial loops are hexagons.

This leads us to parameterize the effective
Hamiltonian in terms
of the various non-trivial AFM loops. 
\begin{equation} \label{eq:heff}
\frac{E_q^{\eff}}{N (\kappa/2)} =
K_0 + K_6 \PP_6 + K_8 \PP_8 + K_{10} \PP_{10} + \cdots \,,
\end{equation}
where $\{K_{2l}\}$ are numerical coefficients, and $\PP_{2l}$ is the number of
non-trivial AFM loops of length $2l$, per site.

To evaluate the coefficients $\{K_{2l}\}$, we need 
to calculate two types of terms:
(i) The number $F(2m)$ of closed paths of total length $2m$ on a decorated
(with trivial $4$-loops)
coordination-$4$ Bethe lattice (Fig.~\ref{fig:paths}e).
(ii) The number $G(2l,2m)$ of closed paths of length $2(m+l)$,
involving a particular
loop of length $2l$ with decorated Bethe lattice paths emanating from each
site along the loop
(Fig.~\ref{fig:paths}f).
Calculating these terms is a matter of tedious but tractable combinatorics.
We find that the functions $F(2m)$, $G(2l,2m)$ 
decay rapidly with $m$, allowing us to sum them in order to 
evaluate the coefficients to any accuracy in Eq.~(\ref{eq:heff}), using
\begin{equation} \label{eq:coeff}
K_0 = \sum_{m=0}^\infty F(2m)  \,, \qquad
K_{2l} =  \sum_{m=0}^\infty G(2l,2m) \,.
\end{equation}
We show the first five coefficients in Tab.~\ref{tab:coeff}.
Thus we have obtained an effective Hamiltonian that is parameterized solely by
the number of AFM loops of various sizes.
Note that the coefficients decay rapidly $K_{2l+2}/K_{2l} \approx 1/10$, 
which leads us to expect short loops to be the dominant terms in the expansion.
This allows us, in principle, to calculate the energy, to any accuracy,
for any member of an 
\emph{infinite} ensemble of classical ground states.
This represents a significant improvement over previous calculations that were
always limited to small system sizes~\cite{sachdev_kagome,bernier}.
\begin{table} 
\begin{tabular}{cll}
\hline
\hline
 coefficient & analytical & numerically fitted \\
\hline
 $K_{0}$ & $-0.59684$ & $-0.59687$ \\
 $K_{6}$ & $-3.482\!\times\!10^{-3}$ & $-3.522\!\times\!10^{-3}$ \\
 $K_{8}$ & $-3.44\!\times\!10^{-4}$ & $-3.76\!\times\!10^{-4}$  \\
 $K_{10}$ & $-3.59\!\times\!10^{-5}$ & $-4.5\!\times\!10^{-5}$ \\
 $K_{12}$ & $-3.8\!\times\!10^{-6}$ & $-5.5\!\times\!10^{-6}$ \\
\hline
\hline
\end{tabular}
\caption{  \label{tab:coeff}
Coefficient values for Eq.~(\ref{eq:heff}), obtained analytically, and by an 
independent numerical fit to the energies in Fig.~\ref{fig:e_LN_fit}.}
\end{table}

Although we derived the effective Hamiltonian for collinear states, 
it turns out that, in fact, the classical tetrahedron zero sum rule implies that
Eq.~(\ref{eq:heff}), with the coefficients in Tab.~\ref{tab:coeff}, is 
valid for \emph{any non-collinear} classical ground state, as well,
with the generalized loop variables expressed as sums over non-trivial loops 
\begin{equation} \label{eq:gen_p}
\PP_{2l}= \frac{1}{\kappa^{2l}}
\sum_{(i_1 \ldots i_{2l})} \mathrm{Re}
(Q^\dagger_{i_1 i_2} Q_{i_2 i_3}
\cdots Q^\dagger_{i_{2l-1} i_{2l}} Q_{i_{2l} i_1})\,.
\end{equation}
Unlike the collinear case, where the elements of $\Qvec^\dagger \Qvec$
could only take the values $0$ or $\kappa^2$, and thus each loop
would contribute $0$ or $1$ to the sum~(\ref{eq:gen_p}), in the general case,
the matrix elements of $\Qvec^\dagger \Qvec$ are complex.

\begin{figure}
\resizebox{!}{5.0cm}{\includegraphics{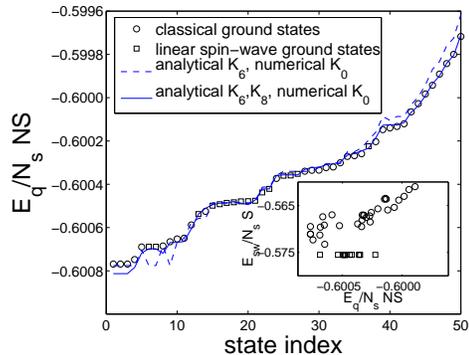}}
\caption{\label{fig:e_LN_fit} 
Calculated energies $E_q$ of 
50 sample classical ground states (open symbols),
16 of which are harmonic spin-wave ground states (squares),
along with $E_q^\eff$, with $2l\!\le\!6$ (dashed line) and 
$2l\!\le\!8$ (solid line).
The constant term $K_0$ was numerically fitted (see main text).
The inset shows the linear spin-wave energy for the same states.
Although the spin-wave energy tends to be lower for states with 
lower $E_q$, the large-$N$ ground state need not be a spin-wave ground state.}
\end{figure}

\emph{Numerical results.}\textemdash
To  verify the validity of the effective Hamiltonian~(\ref{eq:heff}),
we calculated the energy for a large number of collinear classical ground
states, as well as linear spin-wave ground states,
obtained by a random flipping algorithm
described elsewhere~\cite{uh_harmonic}.
We find that the energies are remarkably well described by $E_q^\eff$, 
even when we cut the expansion~(\ref{eq:heff}) off
at $2l=8$, as shown in Fig.~\ref{fig:e_LN_fit}.
We used the coefficient values of Tab.~\ref{tab:coeff}, but had to 
adjust the constant term $K_0$ \emph{separately} for each choice of cutoff,
in order to get a good fit~\cite{footnote}.
In practice, this means that the effective Hamiltonian~(\ref{eq:heff}) is
extremely useful for \emph{comparing} energies of various states,
even with a small cutoff, but requires many terms in order to accurately 
determine the energy.
An independent $5$-parameter numerical fit,
to Eq.~(\ref{eq:heff}), up to $2l=12$, gives
the values shown in the right-hand column of Tab.~\ref{tab:coeff}.

\begin{figure}
\resizebox{!}{4cm}{\includegraphics{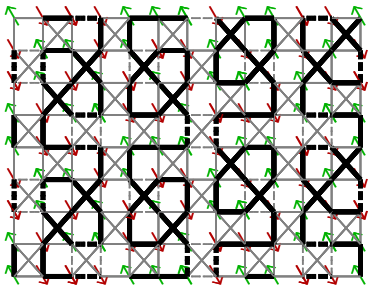}}
\caption{\label{fig:ground_state}
The ground state of our large-$N$ theory,
as viewed in a (001) projection.
Here, light (dark) bonds represent AFM (ferromagnetic) bonds
(unlike in Fig.~\ref{fig:paths}).
The shown pattern is repeated along $x$ and $y$ directions, as well as
in adjacent $z$ slices.
This state has a $48$ site magnetic unit cell.}
\end{figure}

Now that we have an approximate formula for $E_q$, for any collinear classical
ground state, we can systematically search these states,
with large magnetic unit cells, to find a ground state.
We conducted Monte Carlo simulations using a 
Metropolis loop flipping algorithm and the effective energy
of Eq.~(\ref{eq:heff}), for various orthorhombic unit cells
of sizes ranging from $128$ to $3456$ sites, with periodic boundary conditions.
We find a minimum energy of $E_q/(N \kappa/2)=-0.60077 N_s$
for a family of \emph{nearly} degenerate states.
They are composed of layers, that can each be in one of four arrangements, 
resulting in $\sim e^{cL}$ states, where $L$ is the system size,
and $c$ is a constant.
Each of these states has $\PP_6=N_s/3$, which
is the maximum value that we find (but is \emph{not} unique to these states),
and $\PP_8=23N_s/6$.
Upon closer investigation, however, we find that a \emph{unique
ground state} (depicted in Fig.~\ref{fig:ground_state}) is selected.
The energy difference to nearby states is of order $10^{-7}N_s$, corresponding
to the $2l=16$ term.

\emph{Discussion.}\textemdash
It was noted by one of us~\cite{clh_harmonic} (see also \cite{Tcher_HFM})
that, in the pyrochlore, the degeneracy of
ground states of the spin-wave quantum Hamiltonian,
at the lowest order in $1/S$, is 
associated with a \emph{gauge-like} symmetry.
This symmetry characterizes the 
degenerate sub-manifold of collinear spin ground states by the condition
\begin{equation} \label{eq:harmonic_gs}
\prod_{i \in \hexagon} \eta_i = -1 \,,
\end{equation}
for all non-trivial hexagons.
Since the spin-wave theory is expected
to be exact in the limit of infinite $S$, the physical ground state must 
satisfy Eq.~(\ref{eq:harmonic_gs}). 
The state depicted in Fig.~\ref{fig:ground_state}, however, 
\emph{does not}.
Looking at the inset in Fig.~\ref{fig:e_LN_fit}, we find that
states with negative hexagon products tend to have lower large-$N$ energy than 
other states, since they tend to have more AFM loops, 
but this is not a strict rule.
Thus it would seem that the $N\to\infty$, large-$\kappa$ ground state
cannot be the physical ($N=1$) large-$S$ ground state.

\begin{figure}
\resizebox{!}{4.6cm}{\includegraphics{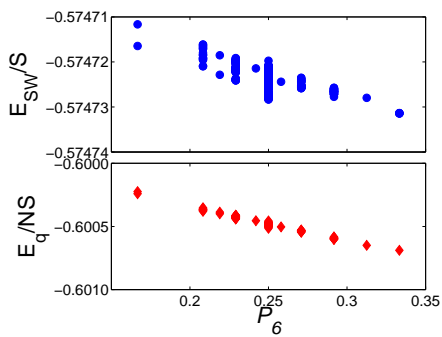}}
\caption{\label{fig:compare_sw}
Per-site large-$N$ energies $E_q$
calculated for various harmonic spin-wave ground states (bottom),
compared to the 
per-site spin-wave energy obtained from an anharmonic calculation for
$S\!=\!1500$ (top).
In both cases the lowest energy is for a state that maximizes the number of
AFM hexagons. }
\end{figure}

Nevertheless, if we restrict the large-$N$
calculation to harmonic spin-wave ground states
only, we find that the ordering of energies for various states is similar
to preliminary anharmonic spin-wave results~\cite{uh_quartic},
and does predict the same ground state.
As shown in Fig.~\ref{fig:compare_sw}, in both cases, the lowest energy
among harmonic spin-wave ground state belongs to a state with the most
AFM hexagons.

The effective Hamiltonian approach that we have outlined here can easily
be applied to other lattices.
In the checkerboard lattice,
the energy is lowest for states that have the most
AFM (square) non-trivial plaquettes.
Thus, the non-degenerate ground state is clearly the
($\pi$,$\pi$) state in which \emph{all} plaquettes are AFM~\cite{bernier}.
In the kagom\'e case, all classical ground states are non-collinear.
However, if we limit ourselves to coplanar arrangements, 
we find that $Q_{ij}$ has the same absolute value for \emph{all}
of the lattice bonds, but the signs differ depending on the \emph{chirality}
of the triangle to which the bond $(i,j)$ belongs. 
Therefore, the effective Hamiltonian~(\ref{eq:heff}), with the generalized
variables~(\ref{eq:gen_p}), prefers classical ground states 
with negative product of triangle chiralities around all hexagons.
One can thus conclude that the ground state is the
$\sqrt{3}\times \sqrt{3}$ state, as large-$N$ calculations have indeed
found~\cite{sachdev_kagome}.
Let us also remark that our method can be generalized to long-range Heisenberg
interactions which are relevant in the context of real materials
like $\mathrm{Tb}_2\mathrm{Ti}_2\mathrm{O}_7$~\cite{ramirez_review,diep}.

Finally, it has been suggested that the disordered (\emph{small}-$\kappa$)
limit of the large-$N$ approximation
for the pyrochlore lattice also has a massive multiplicity of
saddle-points~\cite{tchern_LN}; an effective Hamiltonian
similar to this paper's could organize the handling of this family. 

\acknowledgments

UH and CLH acknowledge support from NSF grant DMR-0240953, 
and PS acknowledges support from the Packard foundation
and the U.S. Dept. of Energy, under Contract No. W-31-109-ENG-38.


\begin{thebibliography}{15}
\bibitem{footnote}
{We could get a cutoff-independent
constant term $\tilde{K}_0$ if we replaced $\PP_{2l}$ by 
$\tilde{\PP}_{2l}= \PP_{2l} - \langle \PP_{2l} \rangle$, where
$ \langle \PP_{2l} \rangle$ is the ensemble average of $\PP_{2l}$, and took
$\tilde{K}_0 = K_0 + \sum K_{2l} \langle \PP_{2l} \rangle$.}.
\end{thebibliography}

\end{document}